\documentstyle[sprocl]{article}

\begin{document}

\title{Proton decay and fermion masses in supersymmetric grand unified
 theories}

\author{Borut Bajc$^{(1)}$, Alejandra Melfo$^{(2,3)}$ and Goran Senjanovi\'c$^{(3)}$\footnote{Invited talk at `` 20 years of SUGRA'', Boston, March 2003.}} 

\address{$^{(1)}$ Jo\v{z}ef Stefan Institute, 1001 Ljubljana, Slovenia
$^{(2)}$CAT, Universidad de Los Andes, M\'erida, Venezuela,$^{(3)}$ICTP, Trieste, Italy. }

\maketitle

\abstracts{We briefly review the issues of proton decay and fermion
masses and mixings in minimal supersymmetric grand unified theories. We
argue that minimal SU(5), although tightly constrained by proton decay
data, is still not ruled out. However, we outline the advantages of SO(10)
unification, in particular in the model with renormalizable see-saw
mechanism and its remarkable predictions of (a) exact R-parity at low 
energies, (b) large atmospheric neutrino angle as a consequence of $b-\tau$ 
unification and (c) 1-3 leptonic mixing angle close to its upper limit.}

\section{Introduction}

Supersymmetric unification  has been the main  extension of the Standard Model
for about 20 years for three main reasons

\begin{enumerate}
\item it provides a natural protection for the weak scale against any 
large scale as long as supersymmetry breaking is around $TeV$ or so
\item it predicts correctly the weak mixing angle \cite{unification}
(it actually anticipated experiment) and makes proton decay
accessible to experiment
\item via the so-called radiative symmetry breaking, it leads
naturally to the Higgs mechanism \cite{Alvarez-Gaume:1983gj}
\end{enumerate} 

Furthermore, in minimal schemes, the enhanced symmetry leads to
relations among quark and lepton masses. The most celebrated one is
$m_b=m_\tau$ at the unification scale \cite{chanowitz}, which is
corroborated by 
experiment once the running is taken into account. This would have
been a great success if not for the fact that the analog relations
fail badly for the second and especially for the first family. 

Another novel feature of supersymmetric unification is the new
source of proton decay through dimension five operators
\cite{w82},  generated by
the superheavy fermionic fields. This operators originate in Yukawa
couplings and are thus intimately correlated with the fermionic mass
and mixing angle relations. In this talk, we review in some detail
this issue in the context of the minimal SU(5) and SO(10) 
supersymmetric GUTs. We will argue that the minimal SU(5) theory is
still in perfect accord with data. However, the non-zero neutrino
masses and the almost perfect conservation of R-parity point naturally
toward its SO(10) extension. We will briefly recall the essential
features of SO(10) grand unification. We emphasize the attractive
feature of R-parity remaining an exact symmetry to all orders in
perturbation theory in the minimal model with a renormalizable see-saw
mechanism. 
We also discuss some interesting new results on fermion
masses such as the deep connection between $b-\tau $ unification and
the large atmospheric neutrino mixing. We conclude with SO(10) being a
complete and realistic model of both unification and fermion masses
and mixings.

\section{The minimal supersymmetric SU(5)}

We define the minimal SU(5) supersymmetric theory in the conventional
manner, that is, with three generations of fermion superfields and a
{\bf 24} adjoint ($\Sigma$) and 5 ($\Phi$) and $\bar 5$ ($\bar \Phi$) Higgs
superfields. This implies the well-known relations $m_D = m_E$
generation by generation, as we mentioned above. Of course, this must
be corrected for the first two generations and in the context of the
minimal model this simply implies the existence of  higher-dimensional
operators cut-off by the Planck scale. Hereafter, by minimal SU(5) we
will mean such a theory, where the higher-dimensional operators are
taken into account. Notice that this is very similar to the neutrino
mass issue in the Standard  Model. The nonvanishing neutrino mass does
not mean that the Standard Model is ruled out, but simply points out
to the existence of  higher-dimensional operators. 
 
 As mentioned above, supersymmetric GUTs in general, and the SU(5)
theory in particular, lead to quite fast proton decay through d=5
operators generated by the superheavy colored triplet Higgs
supermultiplet ($T$ and $\bar T$) with masses $m_T \simeq M_{GUT}$. An
important question in recent years was whether the minimal
supersymmetric SU(5) theory is already ruled out on this basis,
especially after it was found out that the RRRR operators play a
crucial role \cite{Goto:1998qg} 
(in the context of SO(10) this was shown before in
\cite{Lucas:1996bc}), and it was finally argued  that this was indeed true
\cite{Murayama:2001ur}. We will see however that the uncertainty in the masses of
the supermultiplets in $\Sigma$, the lack of knowledge of sfermion and
fermion  masses and mixings, and the presence of higher-dimensional
operators is sufficient to keep SU(5) still in accord with
experiment. 

Let us discuss these points carefully:
\subsection{Determination of $M_{GUT}$ and $m_T$}

The superpotential for the heavy sector is (up to terms $1/M_{Pl}$)

\begin{equation}
W = m Tr\Sigma^2+\lambda Tr\Sigma^3+a{(Tr\Sigma^2)^2\over M_{Pl}}+
b {Tr \Sigma^4\over M_{Pl}}\;.
\end{equation}

Of course, if $\lambda\approx{\cal O}(1)$, we ignore higher-dimensional
terms. However, in supersymmetry $\lambda$ is a Yukawa-type coupling,
i.e. self-renormalizable. For small $\lambda$ ($\lambda\ll M_{GUT}/M_{Pl}$),
the opposite becomes true and $a$ and $b$ terms dominate. In this case,
it is a simple exercise to show that 
\begin{equation}
\label{m34m8}
m_3 = 4 m_8\;,
\end{equation}
\noindent
where $m_3$ and $m_8$ are the masses of the weak triplet and color
octet in $\Sigma$. In the renormalizable case $m_3=m_8$.
The RGE at one loop for the gauge couplings are  readily solved to give:
\begin{eqnarray}
m_T&=&m_T^0\left({m_3\over m_8}\right)^{5/2}\;,\\
M_{GUT}&=&M_{GUT}^0\left({M_{GUT}^0\over 2m_8}\right)^{1/2}\;.
\end{eqnarray}
where the superscript $^0$ denotes the values in
the case $m_3=m_8$.
Since when (\ref{m34m8}) is valid, $m_8\approx M_{GUT}^2/M_{Pl}$, 
we can also write
\begin{equation}
M_{GUT}\approx \left[\left(M_{GUT}^0\right)^3M_{Pl}\right]^{1/4}\;.
\end{equation}
 From (\ref{m34m8}) we get
\begin{equation}
\label{ilia}
m_T=32m_T^0\;\;\;,\;\;\;M_{GUT}\approx 10 M_{GUT}^0\;\;\;.
\end{equation}

Now, $M_{GUT}^0\approx 10^{16}$ GeV and it was shown last year 
\cite{Murayama:2001ur} that $m_T>7\times 10^{16}$ GeV is sufficiently 
large to be in accord with the newest data on proton decay. 
On the other hand, since \cite{Murayama:2001ur}
\begin{equation}
m_T^0<3\times 10^{15}{\rm GeV}\;,
\end{equation}
\noindent
from (\ref{ilia}) we see that $m_3=4m_8$ is enough to save the
theory. Obviously, an improvement of the measurement of $\tau_p$
is badly needed. It is noteworthy that in this case the usual
$d=6$ proton decay becomes out of reach: $\tau_p(d=6)>10^{38}$ yrs.

\subsection{Higher dimensional operators and fermion masses and
mixings}

In the minimal SU(5) theory at the renormalizable level we have the 
Yukawa coupling relations at $M_{GUT}$
\begin{equation}
\label{ymin}
Y_U=Y_U^T\;\;\;,\;\;\;Y_E=Y_D\;\;\;,
\end{equation}
\noindent
where in the supersymmetric standard model language the Yukawa 
sector can be written as 
\begin{eqnarray}
W_Y = && H Q^TY_Uu^c+\bar H Q^T Y_D d^c+\bar H e^{cT}Y_E L
\nonumber\\
&+&{1\over 2}T Q^T\underline A Q + Tu^{cT}\underline B e^c+
\bar T Q^T\underline C L+\bar T u^{cT}\underline D d^c\;.
\end{eqnarray}

Also, in the minimal renormalizable model (at $M_{GUT}$)
\begin{equation}
\label{abcdmin}
\underline A=\underline B=Y_U=Y_U^T\;\;\;,\;\;\;
\underline C=\underline D=Y_D=Y_E\;\;\;.
\end{equation}

The fact that $\underline A=\underline B=Y_U$, 
$\underline C=Y_E$, $\underline D=Y_D$, is simply 
a statement of SU(5) symmetry. On the other hand 
$Y_U = Y_U^T$ and $Y_D=Y_E$ result from the SU(4)$_c$ Pati-Salam (PS)
\cite{Pati:1974yy} like symmetry left unbroken by $\langle H\rangle$ and 
$\langle\bar H\rangle$. Under this symmetry $d^c\leftrightarrow e$, 
$u\leftrightarrow u^c$, $d\leftrightarrow e^c$. It is the above
relations that cause the problem, since the color triplet couplings
are then well-defined.

 Of course, this symmetry is broken 
by $\langle\Sigma_\alpha^\alpha\rangle\ne\langle\Sigma_4^4\rangle$, 
where $\alpha=1,2,3$; this becomes relevant when we include 
higher dimensional operators suppressed by
$\langle\Sigma\rangle/M_{Pl}$, such as 
\begin{equation}
\frac{1}{M_{Pl}} \bar \Phi \, \Sigma \, 10_f \, \bar 5_f
\end{equation}
where $10_f$ and $\bar5_f$ stand for the fermion multiplets. Since
there are now neither SU(5) nor SU(4)$_c$ (PS) symmetries, it is a simple 
exercise to show that the fermion mass relations now become
consistent. For a careful study of proton decay in this context see
\cite{bfs02-1,Emmanuel-Costa:2003pu}, where it is shown that this by 
itself is  enough to save the theory.

\subsection{Sfermion and fermion masses and mixings and proton lifetime}

As we said, d=5 proton decay is generated through Yukawa couplings and
thus fermion and sfermion masses and mixings play an important
role. We shall not discuss this in detail here, since we know nothing
about the sfermion properties; this issue belongs to the realm of
supersymmetry breaking and is orthogonal to grand unification. It is
highly model-dependent, however the constraints from FCNC can be
useful in restricting the proton decay predictions. It has been
discussed extensively in \cite{bfs02-1,bfs02-2} and we refer the
reader to these works. 

In short, the minimal SU(5) supersymmetric theory, although in accord
with experiment is highly constrained and thus potentially
falsifiable. It is the best, most predictable theory of proton decay,
and in our opinion care should be taken before we can proclaim it
dead. Certainly, and order of magnitude or two of improvement in
proton lifetime measurements is more than  welcome.

\section{Supersymmetric SO(10) theory}  

SO(10) grand unified theory is an appealing candidate for the unification 
of quarks and leptons (family by family) and their interactions for a
number of reasons:
\begin{enumerate}
\item a family of fermions, included the right handed neutrino, fits
into a single spinorial representation

\item it offers a natural explanation of the smallness of neutrino
mass through the see-saw mechanism \cite{see-saw} 

\item as any theory of see-saw mechanism, it incorporates
leptogenesis, a simple and appealing mechanism for generating the
baryon asymmetry of the universe \cite{Fukugita:1986hr}

\item left-right symmetry in the form of charge conjugation is a finite
gauge transformation, which allows for spontaneous parity violation in
the context of Left-Right models \cite{Pati:1974yy,lr}

\item it contains the Pati-Salam \cite{Pati:1974yy} 
subgroup which is the prototype of
quark-lepton unification; it is this symmetry that leads to the above
discussed relation $m_D=m_E$.

\item matter parity, which is equivalent to R parity, is also a finite
gauge transformation; furthermore, it remains exact after all the
symmetry breaking has taken place   \cite{amrs99,abmrs01}

\item in the minimal renormalizable see-saw scenario, it can be shown
that SO(10) is a realistic theory of fermion masses and mixings; it
correlates $b-\tau$ unification with a large atmospheric neutrino
mixing \cite{bsv02} (see also \cite{Bajc:2001fe}) and it predicts the element $1-3$ of the leptonic mixing
matrix to be close to the experimental limit \cite{gmn03}
\end{enumerate}
We will now discuss the last two issues in some detail.  

\subsection{SO(10) and R-parity}

It is convenient to work with matter parity defined as $M=
(-1)^{3(B-L)}$ instead of its equivalent analog R-parity, $R =M
(-1)^{2S}$,  since $M$
commutes with supersymmetry. In SO(10), under $M$ 
 \begin{eqnarray}
{\bf 16} \; \stackrel{M}{\longrightarrow} \;  -{\bf 16} \quad , \quad
{\bf 10} \; \stackrel{M}{\longrightarrow} \;  {\bf 10}\;,
\label{mparity}
\end{eqnarray}
 which is equal to $C^2$, where $C$ is the center of SO(10). Of
course, the crucial thing is the gauging of $B-L$ \cite{b-lgauging}. Assume
next that one utilizes {\bf 126}-dimensional representation in order to
break $B-L$ symmetry and give a mass to the right-handed
neutrino, just as in the original  proposal. This allows us
to stick to the renormalizable theory and simultaneously to keep
R-parity intact. The alternative, to use a 16-dimensional multiplet 
\cite{so10} in
order to break $B-L$,  breaks R-parity at the GUT scale, and the theory
must thus be augmented by extra discrete symmetries in order to keep
the proton stable. 
This would take us beyond the SO(10) theory which we pursue here. 

This however is not sufficient. Both right-handed and left-handed
sneutrino fields can in principle get nonvanishing VEVs and break
R-parity. As far as the right-handed sneutrino is concerned, it is a
simple exercise to show that the group structure does not allow
it \cite{amrs99}. Since its mass is equal to the mass of the right-handed neutrino,
and thus very large (close to GUT scale), this fact survives a tiny
supersymmetry breaking. 

The situation with the left-handed neutrino is far more complex. It
belongs to the light sector of the theory and a priori it could be
tachyonic to start with, and thus get a VEV. To see that this does not
happen requires a subtle interplay between cosmology, phenomenology
and the decoupling theorem. First of all, recall that the left-handed
sneutrino cannot be a tachyon in  the MSSM with R-parity. This we know
from the Z-decay width; namely, the opposite would result in the
existence of a Goldstone boson, the Majoron ($J$), and the Z-boson would
decay into a Majoron and its real counterpart ($R_J$). The mass of
$R_J$ is of the order of the sneutrino VEV, and this VEV must be small
in order for the Majoron not to get overproduced in stars
\cite{Georgi:1981pg} and/or for
the induced neutrino mass to be small enough. Therefore the
left-handed sneutrino VEV would produce a Z decay width incompatible
with observations.

A simple application of the decoupling theorem ensures that the same
happens in SO(10) theory. In this case a would-have-been Majoron of
course gets a mass, but this mass must vanish in the limit of large
$M_R$, the scale of $B-L$ breaking. More precisely, one
shows that \cite{amrs99}
\begin{equation}
m_J \simeq \frac{m_{3/2}^2}{M_R}
\end{equation}
where $m_{3/2}$ is the scale of low-energy supersymmetry
breaking. Obviously $m_J \ll M_Z$, and the above reasoning follows
through. Remarkably enough, R-parity can be not  broken at all
in SO(10), independently of the mechanism of supersymmetry breaking
 \cite{abmrs01}.

This important fact holds true for any renormalizable theory of the
supersymmetric see-saw mechanism based on spontaneous breaking of
$B-L$ \cite{amrs99,abmrs01,ms03}. This not only guarantees a stable enough proton, but
also a stable LSP (lightest supersymmetric particle), probably the
most natural candidate for the dark matter of the Universe.

\subsection{SO(10) and fermion masses and mixings}

The minimal SO(10) theory, by definition, contains a single
10-dimensional Higgs supermultiplet coupled to fermions. At first
glance, in the usual manner, this should imply $m_D = m_E$ as stressed
repeatedly. Namely, the Higgs that takes the VEV
is a bi-doublet
(2,2,1) under the Pati-Salam group $SU(2)_L\times SU(2)_R\times
SU(4)_c$. However, the {\bf 126} representation needed for the see-saw
mechanism contains also a (2,2,15) field, and in general this field
gets a VEV \cite{Babu:1992ia} .
 In the supersymmetric theory, this amounts to using a {\bf 210}
Higgs multiplet at the GUT scale. This has a number of interesting
features and should be pursued and studied more carefully. 

For us, it is sufficient to have a (2,2,15) with a VEV. This simple
fact, surprisingly enough, renders the theory completely realistic as
we shall see. On
top of that it offers a natural connection between the large
$\mu-\tau$ mixing and $b-\tau$ unification. This provides a badly
needed answer as to why the small quark mixing angle(s) should be
accompanied by large leptonic mixing(s). Let us see how this comes
about.

First of all, let us recall an important fact in the SO(10) see-saw
mechanism.
The Yukawa couplings are given by
\begin{equation}
{\cal L}_Y=10_H\psi Y_{10}\psi+126_H\psi Y_{126}\psi\;,
\end{equation}
where $\psi$ stands for the $16$ dimensional spinors which 
incorporate a family of fermions, and $Y_{10}$ and $Y_{126}$ 
are the Yukawa coupling matrices in generation space. 

From
\begin{equation}
126_H=(3,1,10)+(1,3,\overline{10})+(2,2,15)+(1,1,6)
\end{equation}
\noindent
one has 
\begin{equation}
M_{\nu_R}=Y_{126}\langle (1,3,\overline{10})_{126}\rangle\;,
\end{equation}
\noindent
where $\langle(1,3,\overline{10})_{126}\rangle =M_R$, the scale of 
SU(2)$_R$ gauge symmetry breaking. 

It can be shown that, after the SU(2)$\times$U(1) breaking through 
$\langle 10_H\rangle =\langle (2,2,1)\rangle\approx M_W$, the 
$(3,1,10)$ multiplet from $126_H$ gets a small vev 
\cite{Mohapatra:1980yp,Magg:1980ut}
\begin{equation}
\langle (3,1,10)_{126}\rangle\propto{M_W^2\over M_P}\;,
\end{equation}
\noindent
where $M_P$ is the scale of the breakdown of parity. In turn, 
neutrinos pick up small masses 
\begin{equation}
\label{seesaw}
M_{\nu_L}=Y_{126}\langle (3,1,10)_{126}\rangle +
m_D^TM_{\nu_R}^{-1}m_D\;,
\end{equation}
\noindent
where $m_D$ is the neutrino Dirac mass matrix. 
It is often assumed, for no reason whatsoever, that the second term 
dominates. This is the so-called type I (or canonical) see-saw \cite{typeI}. 
In what follows we explore the opposite case,type II (or
non-canonical) see-saw. 
After all, it does not involve Dirac mass terms and so there is no 
reason a priori in this case to expect quark-lepton analogy of 
mixing angles. In this sense the type II see-saw is 
physically more appealing. 
More than that, we will show that the large leptonic 
mixing fits perfectly with the small quark mixing, as long as 
$m_b=m_\tau$. 

To see this, notice that fermion masses take the following form
\begin{eqnarray}
\label{mu}
M_U&=&Y_{10}v_{10}^u+Y_{126}v_{126}^u\;,\\
M_D&=&Y_{10}v_{10}^d+Y_{126}v_{126}^d\;,\\
M_E&=&Y_{10}v_{10}^d-3Y_{126}v_{126}^d\;,\\
\label{mn}
M_N&=&Y_{126}\langle (3,1,10)_{126}\rangle\;,
\end{eqnarray}
\noindent
where $U$, $D$, $E$, $N$ stand for up quark, down quark, 
charged lepton and neutrino, respectively, while 
$v_{10}^{u,d}$ and $v_{126}^{u,d}$ are the two 
vevs of $(2,2,1)$ in $10_H$ and $(2,2,15)$ in $126_H$, 
and the last formula is the assumption of the 
type II see-saw. The result is surprisingly simple. 
Notice that \cite{Brahmachari:1997cq}
\begin{equation}
\label{mnde}
M_N\propto Y_{126}\propto M_D-M_E\;.
\end{equation}

Now, let us study the $2^{nd}$ and $3^{rd}$ generations, 
and work in the basis of $M_E$ diagonal. The puzzle then is: 
why a small mixing in $M_D$ corresponds to a large mixing in 
$M_N$? For simplicity take the mixing in $M_D$ 
to vanish, $\theta_D=0$, and ignore the second generation masses, 
i.e. take $m_s=m_\mu=0$. Then
\begin{eqnarray}
\label{mnu}
M_N\propto\pmatrix{
  0 
& 0
\cr
  0
& m_b-m_\tau
\cr}\;.
\end{eqnarray}
Obviously, unless $m_b=m_\tau$, neutrino mixing vanishes. Thus, 
large mixing in $M_N$ (the physical leptonic mixing in the 
above basis) is deeply connected with the $b-\tau$
unification. Notice that we have done no model building whatsoever; 
we only assumed a renormalizable SO(10) theory and the type II
see-saw. 

Of course, this is only very qualitative; for a more careful analysis,
one needs to switch on $m_\mu$, 
$m_s$ and the mixings. This has been done in detail in \cite{bsv02}, and we
refer the reader there, suffices it to say that the precise
computations confirm the above qualitative analysis.

This is all nice, you may say, but what about three generations? The
analysis has just been performed \cite{gmn03}, and it will be reported in
the talk by Rabi Mohapatra in this conference. We only cite here the
crucial prediction of the theory: the 1-3 leptonic mixing is as large
as the experiments allows, making the theory all the more
interesting.

\section{Conclusions}

In these talk, we have focused on some of the central issues in
supersymmetric grand unification: the predictions for proton decay
and its natural connection with fermion masses and mixings. We started
off by showing that the minimal supersymmetric SU(5) theory, although
still alive, is tightly constrained and can and should be tested by
improved proton decay experiments. 

To us, however, SO(10) appears a more complete and more appealing
theory. First of all, it succeeds where the MSSM and SU(5) fail, that
is, in determining the low energy effective theory. We now know that
in the context of SO(10), at least with a renormalizable see-saw
mechanism, R-parity can never be broken and neutralinos ought to be
some or all of the dark matter of the universe. It is reassuring that
the same theory provides a complete and consistent description of all
fermion masses and mixings. Most important, if you stick to the
type II see-saw the $b-\tau$ unification says simply and clearly that
the small quark $u-b$ mixing must be accompanied by a large
atmospheric neutrino mixing. Last but not least, it seems that the 1-3
neutrino mixing will provide a crucial test of the theory in the near
future. 

In spite of the apparent success, a lot remains to be done. The most
urgent and most important task is a careful study of proton
decay. Namely, the possible existence of intermediate scales in SO(10)
lowers the unification scale, and thus the proton lifetime. This is
more than welcome for d=6 operators mediated proton decay, since it
makes it accessible to 
experiment. However dimension 5 operators run the risk of leading to a
too fast decay. We have seen that in SU(5) they are safe, but only
marginally.

\vspace{0.5cm}

\newpage
\noindent {\em \large Acknowledgements}

We are deeply grateful to our collaborators
 Charanjit  Aulakh,
Pavel Fileviez-P\'erez, Andrija Ra\v{s}in and Francesco ``Paco''
Vissani.

\end{document}